 \documentclass[final,5p,twocolumn]{elsarticle}

\makeatletter
\def\ps@pprintTitle{%
 \let\@oddhead\@empty
 \let\@evenhead\@empty
 \def\@oddfoot{}%
 \let\@evenfoot\@oddfoot}
\makeatother

\usepackage{amsmath}

\usepackage{algorithm}
\usepackage{algorithmicx}
\usepackage[noend]{algpseudocode}
\floatname{algorithm}{Algorithm}

\usepackage{amssymb}
\usepackage{comment}
\usepackage{eucal}
\usepackage{amsfonts}
\usepackage{enumerate}
\usepackage{graphicx}
\usepackage{setspace}
\usepackage{epstopdf}
\usepackage{subfigure}
\usepackage{amsmath}
\usepackage{array}
\usepackage{color}
\usepackage{comment}
\usepackage{tikz}
\usepackage{amsthm}

\newcommand*\circled[1]{\tikz[baseline=(char.base)]{\node[shape=circle,draw,inner sep=0.6pt] (char) {#1};}}

\newcommand{\ie}{{\em i.e.}}
\newcommand{\eg}{{\em e.g.}}
\newcommand{\et}{{\em et al.}}

\newtheorem{theorem}{Theorem}
\newtheorem*{theorem*}{Theorem}
\newtheorem*{proof*}{Proof}

\newtheorem{definition}{Definition}

\newcommand*{\QEDA}{\hfill\ensuremath{\blacksquare}}%

\journal{Journal of Network and Computer Applications}

\begin{document}

\begin{frontmatter}



\title{ERA: Towards Privacy Preservation and Verifiability for Online Ad Exchanges}


\author{Chaoyue~Niu}
 \ead{rvincency@gmail.com}

\author{Minping~Zhou}
 \ead{zhouminping1991@gmail.com}

 \author{Zhenzhe~Zheng}
 \ead{zhengzhenzhe220@gmail.com}

 \author{Fan~Wu\corref{cor1}}
 \ead{fwu@cs.sjtu.edu.cn}

 \author{Guihai~Chen\corref{cor2}}
 \ead{gchen@cs.sjtu.edu.cn}

\address{Shanghai Key Laboratory Scalable Computing and Systems \\ Shanghai Jiao Tong University, China}

\cortext[cor1]{F. Wu is the corresponding author.}

\tnotetext[t1]{This work was supported in part by the State Key Development Program for Basic Research of China (973 project 2014CB340303), in part by China NSF grant 61672348, 61672353, 61422208, and 61472252, in part by Shanghai Science and Technology fund 15220721300, and in part by the Scientific Research Foundation for the Returned Overseas Chinese Scholars. The opinions, findings, conclusions, and recommendations expressed in this paper are those of the authors and do not necessarily reflect the views of the funding agencies or the government.}
\tnotetext[t2]{This manuscript is the author's version of the paper~\cite{jour:jnca:NiuZZWC17} accepted by Journal of Network and Computer Applications, and the full version of the paper~\cite{proc:globecom15:zhou} accepted by IEEE Globecom 2015.}

\begin{abstract}
Ad exchanges are kind of the most popular online advertising marketplaces for trading ad spaces over the Internet. Ad exchanges run auctions to sell diverse ad spaces on the publishers' web-pages to advertisers, who want to display ads on ad spaces. However, the parties in an ad auction cannot verify whether the auction is carried out correctly or not. Furthermore, the advertisers are usually unwilling to reveal their sensitive bids and identities. In this paper, we jointly consider the auction verifiability and the advertisers' privacy preservation, and thus propose ERA, which is an \underline{E}fficient, p\underline{R}ivacy-preserving, and verifi\underline{A}ble online auction mechanism for ad exchanges. ERA exploits an order preserving encryption scheme to guarantee privacy preservation, and achieves verifiability by constructing a novel protocol of privacy preserving integer comparison, which is built on the Paillier homomorphic encryption scheme. We extensively evaluate the performance of ERA, and our evaluation results show that ERA satisfies several desirable properties with low computation, communication, and storage overheads, so ERA can be easily deployed in today's ad exchanges.
\end{abstract}

\begin{keyword}


Ad Exchange; Privacy Preservation; Verifiability; Auction Theory.

\end{keyword}

\end{frontmatter}


\section{Introduction}\label{sec:intro}
Ad exchange is considered as a new type of digital marketplace, where ad places on the web-pages can be traded in real-time via an auction mechanism~\cite{Muthukrishnan:2009:AER:1697167.1697169,mansour2012doubleclick}. A number of ad exchanges have emerged on the Internet, such as DoubleClick~\cite{DoubleClick}, Adecn~\cite{Adecn}, RightMedia~\cite{Rightmedia}, and OpenX~\cite{OpenX}. Ad exchanges serve as a highly effective and efficient tool for selling and buying advertisements, and benefit both publishers and advertisers~\cite{balseiro2014yield,dvovrak2014online,schain2013ad}.
There are billions of ad transactions per day across more than $2$ million websites~\cite{GoogleWhitePaper}, and Internet companies, such as Google, Microsoft, and Yahoo, have extracted a large amount of revenue every year from the ad transactions in their ad exchange platforms~\cite{Muthukrishnan:2009:AER:1697167.1697169}.


In ad exchanges, second-price auction, promoting the ideal of competitive pricing and economic efficiency~\cite{JNCA2016:Auction}, is regarded as the most important technique to allocate ad spaces~\cite{ben2015ad,balseiro2015repeated}. An ad auction allows each interested advertiser to bid for a certain ad space, and the highest bidding advertiser gets the opportunity to display her/his advertisement. However, the current ad auctions have two critical security problems, \ie, privacy leakage and auction manipulation. On one hand, each advertiser needs to submit her/his sensitive identity and bid to participate in the ad auction, which will inevitably breach her/his privacies. Moreover, such type of privacy leakage brings in bidding unfairness, since the advertiser, who knows other advertisers' bidding strategies, possibly gains a huge advantage in the present or subsequent auctions~\cite{parkes2008practical}. On the other hand, publishers and advertisers have no control over the outcome determination, and are forced to unconditionally accept it, even if the ad exchange may manipulate or corrupt the auction~\cite{bidrigging}. Under such paradigm, the correctness of ad auctions is totally relied on the reputations of ad exchanges. Therefore, it is highly essential to design a privacy preserving and verifiable auction mechanism, where ad exchanges are able to calculate the auction outcome, and prove its correctness without leaking the private information of advertisers. If {privacy preservation} and {verifiability} can be guaranteed simultaneously, ad exchanges will certainly attract a larger scale of advertisers and publishers to engage in.





Unfortunately, existing auction mechanisms rarely considered these two properties at the same time. For example, the auction mechanisms in~\cite{parkes2008practical,RMMY12,angel2013verifiable} achieved {verifiability} under the assumption that the bidding information should be revealed to the auctioneer. In contrast, some researchers have proposed potential solutions for bid protection~\cite{6566870,huang2015pps,privacy15INFOCOM}, but they ignored the consideration of verifiability. Furthermore, the auction in the online ad exchange significantly differs from these conventional auction mechanisms due to the following two features:
\begin{itemize}
  \item \textbf{Low Latency}: Unlike traditional goods, ad spaces are extremely perishable. If an auction for a single ad space does not complete before the web-page is loaded into the user's browser, then the opportunity to place an ad is lost. In particular, the time for executing an ad auction is usually limited in a short interval, \eg, typically 100 milliseconds~\cite{DAERTBP}.
  \item \textbf{Large Scale}: There are billions of ad auctions per day with millions of advertisers participating in~\cite{mansour2012doubleclick}.
\end{itemize}



Considering above two requirements, there are three major challenges to integrate privacy preservation and verifiability in ad exchanges. The first and the thorniest design challenge comes from the efficiency requirement of ad auctions, \ie, ad auctions should support a large scale of advertisers with low latency. Specifically, the online phase of winner determination and payment calculation should be evaluated within 100ms, which in turn hinders the application of heavy-weight cryptosystems even if they provide strong security assurance. Yet, another challenge is that auction execution and privacy preservation seem to be contradictory objectives. Evaluating the outcome of an ad auction allows the auctioneer to examine all the bids to determine the winner and corresponding payment, whereas preserving the privacy tends to prevent her/him from learning these confidential contents. The last but not least design challenge is how to guarantee verifiability. Although the auction verification can be performed offline, and has no strict time constraint, it is never an easy job to design a verifiable protocol without breaking the property of privacy preservation.

In this paper, we jointly consider the three design challenges, and thus propose ERA, which is an \underline{E}fficient, p\underline{R}ivacy-preserving, and verifi\underline{A}ble online auction mechanism for ad exchanges. ERA first formalizes the allocation of ad spaces into a three-tier auction model, including one auctioneer, ad networks, and advertisers, where intermediary ad networks can help the auctioneer calculate the auction outcome in a parallel way. ERA then employs an order preserving encryption scheme to generate a set of mapped bids, which hide the exact values of original bids while maintaining their ranking order. Therefore, the auction can be executed in the ciphertext space to guarantee bid protection. Besides, for identity preservation, each ad network needs to mask her/his bidder members' unique identities. At last, to facilitate verification, ERA constructs a privacy preserving integer comparison protocol by capitalizing the homomorphic properties of the Paillier cryptosystem.




We summarize our key contributions as follows:
\begin{itemize}
\item We first model the problem of ad space allocation in ad exchanges as a second-price auction model, in which there are one auctioneer, a set of intermediary ad networks, and their subscribed ad bidders.
\item To the best of our knowledge, ERA is the first online auction mechanism achieving both privacy preservation and verifiability. Besides, specific to the ad exchange, we incorporate the considerations of its three-tire structure and stringent efficiency requirement.
\item We have implemented ERA, and extensively evaluate its performance. Our evaluation results show that ERA achieves good effectiveness and efficiency in the large-scale ad exchanges. In particular, when the auction latency is limited within 10ms, ERA with 100 ad networks can support more than 5 million advertisers.
\end{itemize}

The remainder of this paper is organized as follows. In Section~\ref{sec:pre}, we introduce the ad exchange model, and present the security requirements. We describe some relevant cryptographic techniques in Section~\ref{sec:crypto}, and propose a protocol of privacy preserving integer comparison in Section~\ref{sec:ppic}. The detailed design of ERA is presented in Section~\ref{sec:ERA}. In Section~\ref{sec:eval}, we evaluate ERA, and show the evaluation results. In Section~\ref{sec:related:work}, we briefly review the related work. Finally, we draw conclusions and give our future work in Section~\ref{sec:conclusion}.

\section{Preliminaries}\label{sec:pre}

\begin{table}[t]
	\caption{\textsc{Notations}} \label{tab:Notation}
	\centering
	\resizebox{\columnwidth}{!}{
		\begin{tabular}[t]{l|p{9.4cm}}
			\hline
			Notation&Remark \\
			\hline\hline
			$\Theta =\{\theta_1,\theta_2,\ldots, \theta_z\}$ & The original bid space consisting of $z$ possible bids\\
			$\widehat{\Theta} =\{\hat{\theta}_1,\hat{\theta}_2,\ldots, \hat{\theta}_z\}$ & The mapped bid space generated via the order preserving encryption scheme\\
			$\mathbb{S} = \{s_1, s_2, \ldots, s_l\}$ & The unique identities of $l$ ad bidders\\
			$\mathbb{B} = \{b_1, b_2, \ldots, b_l\}$ & The original bids of $l$ ad bidders\\
			$\hat{\mathbb{B}}=\{\hat{b}_1,\hat{b}_2,\ldots,\hat{b}_l\}$ & The mapped bids of $l$ ad bidders\\
			$\mathbb{A} = \{a_1, a_2, \ldots, a_w\}$ & The set of $w$ ad networks\\
			$s_{max}, b_{sec}$ & The winner and corresponding payment\\
			$G_\rho$ & The multiplicative cyclic group of prime order $\rho$\\
			$g, h$ & Two generators of $G_\rho$\\
			$n, \phi$& Public and private keys of the Paillier cryptosystem\\
			$E_n(\cdot), D(\cdot)$ & Paillier encryption and decryption algorithms\\
			\hline
		\end{tabular}
	}
\end{table}

In this section, we first describe system and auction models for ad exchanges. We then present two desirable security requirements on the design. The frequently used notations are listed in Table~\ref{tab:Notation}.


\subsection{System Model}\label{sec:system:model}
We consider a real-time online advertising marketplace, where there are web users, publishers, one ad exchange, ad networks, and advertisers. Generally, every view of web users on a publisher's web-page stimulates the conduction of a second-price ad auction, in which the ad spaces on the web-page are efficiently allocated among advertisers. We now present the system model of an ad exchange, which is based on the model proposed by Muthukrishnan~\cite{Muthukrishnan:2009:AER:1697167.1697169}, and is a generation of current ad exchange models in the literature~\cite{DoubleClick}.

\begin{figure}[t]
\centering
\includegraphics[width=0.95\columnwidth]{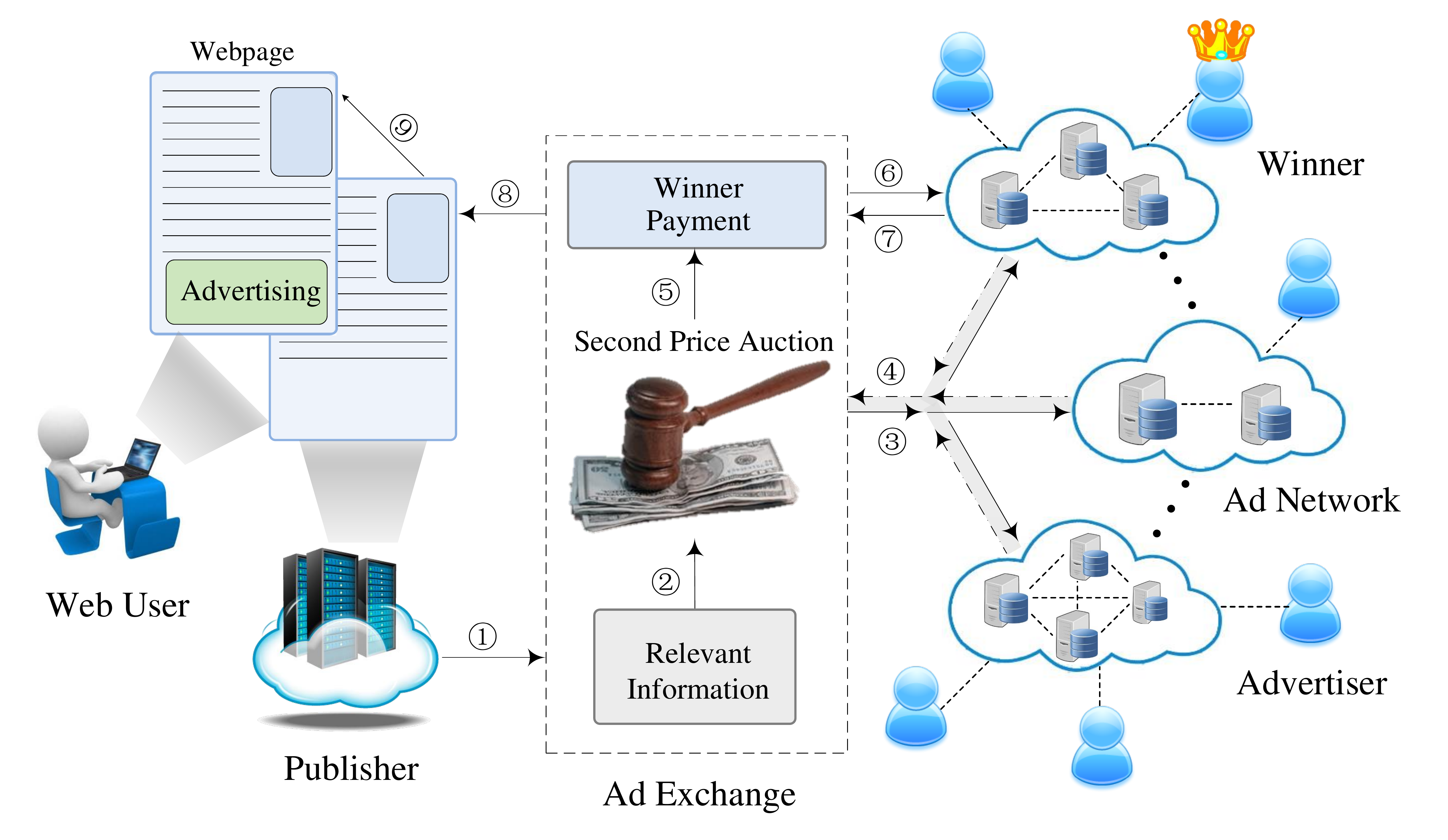}
\caption{An ad exchange ecosystem.}\label{fig:1}
\end{figure}

As shown in Figure~\ref{fig:1}, an ad auction is initiated when a web-user visits to a publisher's web-page, which contains a HTML iframe of JavaScript snippet that generates an ad request to the ad exchange (Step \circled{1}). From the ad request, the ad exchange can extract the relevant information, \eg, the behaviour features of the web-user, the reserve price set by the publisher, the type of the ad space, and the time stamp (Step \circled{2}). The relevant information is delivered to all advertisers through ad networks (Step \circled{3}). Based on the provided information, advertisers can accurately derive the valuation over the ad space. The advertisers submit their bids, which are calculated according to their valuations, to the ad networks that they belong to, and then to the ad exchange (Step \circled{4}). According to the reported bids, the ad exchange runs a second-price auction to determine the winner and her/his payment, which are published to the public (Step \circled{5}). The ad exchange requires the winner to submit her/his ad tag and charge, both of which are further sent to the publisher (Step \circled{6}, \circled{7}, and \circled{8}). In the end, the publisher displays the winner's advertisement on her/his web-page (Step \circled{9}). The whole process typically completes within 100ms.

\subsection{Auction Model}\label{s2_2}

In this subsection, we model the ad space allocation as a sealed bid auction with a single item. The trading items in ad auctions are ad spaces, which can be classified into several types based on the information of web-pages, \eg, videos, images, and texts. Without loss of generality, we consider a specific type of ad space in the following discussion. In our ad auction model, there are four major entities: ad bidders, ad networks, an agent, and an auctioneer, which are introduced in details as follows.

\emph{Ad bidders:} The unique identities of $l$ ad bidders are denoted by $\mathbb{S}=\{s_1,s_2,\ldots,s_l\}$. Each bidder $s_i \in \mathbb{S}$ has an \emph{original} bid $b_i$ for the trading ad space, and an ad tag indicating the identity of her/his advertisement. The original bids of all bidders are denoted by $\mathbb{B} = \{b_1, b_2, \ldots, b_l\}$.

\emph{Agent:} We introduce a new party, called agent, to provide bidders with \emph{mapped} bids, which are used to design a privacy preserving auction. The agent employs an order preserving encryption scheme to construct the set of mapped bids, denoted by $\hat{\mathbb{B}}=\left\{\hat{b}_1,\hat{b}_2,\ldots,\hat{b}_l \right\}$. We assume that the agent is honest-but-curious\footnote{The agent is honest-but-curious if she follows the designed protocol but tries to gather information about other participants~\cite{ccs16:OT}.} in our adversary model.

\emph{Ad networks:} The set of $w$ ad networks is denoted by $\mathbb{A}=\{a_1,a_2, \ldots, a_w\}$. Each ad network $a_j \in \mathbb{A}$ manages a few subscribed ad bidders, who pre-store their mapped bids and ad tags on the ad network $a_j$. In this way, the bidders do not need to encrypt and submit their bids on the fly, and thus the interaction between bidders and ad networks can be significantly reduced.

\emph{Auctioneer:} The auctioneer, acting as the ad exchange, calculates the auction outcome: the winner and her/his corresponding payment.

The ad auction is divided into two stages: the internal auction stage for the ad bidders in each ad network, and the global auction stage for all ad bidders. We employ the classical second-price auction to determine the outcome of each auction stage. Specifically, in the internal auction stage, each ad network $a_j \in \mathbb{A}$ selects the highest and second highest mapped bids from her/his subscribed ad bidders, and then sends to the auctioneer. In the global auction stage, after collecting all the pairs of the highest and second highest mapped bids from ad networks, the auctioneer then chooses the globally highest and second highest mapped bids, denoted as $\hat{b}_{max}$ and $\hat{b}_{sec}$. We also let $s_{max}$ denote the identity of the bidder with the globally highest bid. Besides, the original bids of $\hat{b}_{max}$ and $\hat{b}_{sec}$ are $b_{max}$ and $b_{sec}$, respectively. The outcome of an ad auction can be expressed as: the winner $s_{max}$ and her/his payment ${b}_{sec}$.


In practical ad auctions, on one hand, each bidder wishes to keep her/his bid and identity secret so that other bidders do not gain an advantage over her/him in the present or subsequent auctions. On the other hand, ad bidders and any system participant want to be able to verify whether the result of an auction is correct, even in the face of a corrupt auctioneer. Thus, we are interested in designing secure ad auctions with the following properties simultaneously.


\begin{definition}[Privacy Preserving Ad Auction~\cite{privacy15INFOCOM,huang2015general}]
An ad auction is privacy preserving if it satisfies:
\begin{enumerate}[(1)]
\item Bid protection: the original bid of each ad bidder should be hidden from all the ad networks, the auctioneer, the agent, and the other ad bidders;
\item Identity preservation: each ad bidder's unique identity can only be revealed to her/his ad network.
\end{enumerate}
\end{definition}

\begin{definition}[Verifiable Ad Auction~\cite{parkes2008practical,angel2013verifiable}]
An ad auction is verifiable if the correctness of outcome can be verified by ad bidders and any third party.
\end{definition}



\section{Cryptographic Tools}\label{sec:crypto}
In this section, we briefly review some relevant cryptographic techniques. We first present a secure and efficient information platform, called certificated bulletin board. We then introduce order preserving encryption, oblivious transfer, and Paillier homomorphic encryption, which are building blocks of our following design.

\subsection{Certificated Bulletin Board}
Certificated bulletin board is an electronic version of traditional bulletin board~\cite{eurocrypt97:bulletin,proc:icde17:niu}, which can be a public and trustworthy website maintained and updated by a certain authority, \eg, the auctioneer in our model. A certificated bulletin board can be read by anybody, but can be written only by some authorized parties, such as the auctioneer and the ad networks. For non-reputation, all posts on the certificated bulletin board should be digitally signed ahead of publication. We introduce such a public broadcast channel to solve the problem of information asymmetry among different system participants, and thus to facilitate the auction verification.


\subsection{Order Preserving Encryption Scheme}
Order preserving encryption scheme, introduced by Agrawal~\et~\cite{Agrawal:2004:OPE:1007568.1007632}, is a representative technique that preserves the ordering of plaintexts in the ciphertext space. In particular, given a set of numeric data $\mathbb{M} = \{m_1,m_2,\ldots, m_z\}$ with $m_1 \geq m_2 \geq \ldots \geq m_z$, the function $OPES(\cdot)$ maps each original data $m_i$ to the mapped data $\hat{m}_i = OPES(m_i)$, such that $\hat{m}_1 \geq \hat{m}_2 \geq \ldots \geq \hat{m}_z$.

Order preserving encryption enables the comparison operation to be directly applied on the encrypted data without decrypting them. By exploiting this property, the outcome of the ad auction can be evaluated among the mapped bids, and thus the bid protection can be achieved.

\subsection{Oblivious Transfer}
\begin{algorithm}[!t]                      
\caption{1-out-of-z Oblivious Transfer ($OT_z^1$)}          
\label{alg:OT}                           
\begin{algorithmic}[1]
\Ensure{}
\Require{\textbf{System parameters}: $\left(g,h,G_\rho\right)$;}
\Require{\textbf{Sender's input}: $\mathbb{M}  = (m_1,m_2,\ldots,m_z)$;}
\Require{\textbf{Receiver's choice}: $\alpha$;}\\
Receiver sends: $y=g^rh^\alpha$, $r \in_R Z_\rho$;\\
Sender replies $\xi_i=(g^{k_i},m_i(y/h^i)^{k_i}), k_i \in_R Z_\rho, 1 \leq i \leq z$;\\
By $\xi_\alpha=(a,b)$, receiver computes $m_\alpha=b/a^r$.
\end{algorithmic}
\end{algorithm}

Oblivious transfer~\cite{1261831} describes a two-party paradigm of secret exchange between a sender and a receiver. In particular, the sender has $z$ messages $\mathbb{M} = \{m_1,m_2,\ldots,m_z\}$, and the receiver wants to know one of the messages, \eg, $m_\alpha$. The oblivious transfer scheme guarantees that the receiver just obtains the message $m_\alpha$ without knowing the other $z-1$ messages, while the sender has no idea of the receiver's choice $\alpha$. Algorithm~\ref{alg:OT} shows the pseudocode of 1-out-of-z oblivious transfer $OT^1_z$, where the multiplicative cyclic group $G_\rho$ is of the prime order $\rho$, and $g, h$ are its two generators.

The oblivious transfer scheme allows each bidder to obtain a mapped bid from the agent, while not disclosing her/his original bid.

\subsection{Paillier Homomorphic Encryption Scheme}
To facilitate the practical computation on encrypted data, Paillier~\cite{PP99} introduces a somewhat homomorphic encryption scheme, which preserves the group homomorphism of addition and allows multiplication by a constant. As shown in Algorithm~\ref{alg:PHES}, $p$ and $q$ are two large primes. The parameter $r$ is a random value chosen from $\mathbb{Z}_n^*$, \ie, $\textrm{gcd}(r, n) = 1$. Besides, we recall that the private key $\phi$ is actually the Euler's totient function of $n$, the number of integers relatively prime to $n$. Moreover, the ciphertext of a message $m$ encrypted under the public key $n$ using the random value $r$ is denoted as $C=E_n(m,r)$. Sometimes, $r$ is omitted to simplify the notation as $C=E_n(m)$.

\begin{algorithm}[!t]                      
\caption{Paillier Homomorphic Encryption Scheme}          
\label{alg:PHES}                           
\begin{algorithmic}[0]
\\
\textbf{System parameters}: $(p, q, r)$\\
\textbf{Public key}: $n = pq$\\
\textbf{Private key}: $\phi =(p-1)(q-1)$\\
\textbf{Encryption}: $C=E_n(m,r)=(1+mn)\cdot r^n(mod~n^2)$;

$\quad\quad\quad\quad\ C^{-1}=E_n^{-1}(m,r)=(1-mn)\cdot r^n(mod~n^2).$\\
\textbf{Decryption}: $m=D(C,\phi)=\frac{(C^\phi-1)/\phi~mod~n^2}{n}$;

$\quad\quad\quad\quad\ m = D(C,r) = \frac{(C\cdot r^{-n}~mod~n^2)-1}{n}$.\\
\textbf{Random Value Recovery}: $r = C^{n^{-1} (mod ~ \phi)} (mod~n).$
\end{algorithmic}
\end{algorithm}





The Paillier scheme has three desirable properties as follows:
\begin{itemize}
  \item \emph{Information Theoretic Hiding}: Due to intractability of Decisional Composite Residuosity Assumption (DCRA)~\cite{PP99}, it is hard for any probabilistic polynomial-time adversary to recover the plaintexts from the ciphertexts without knowing the private key or the random value.
  \item \emph{Computationally Binding}: Given two different pairs of plaintext and random value $\left\{(m_1, r_1), (m_2, r_2)|m_1 \neq m_2, r_1 \neq r_2\right\}$, it is infeasible to encrypt $m_1$ and $m_2$ such that $E_n(m_1,r_1) = E_n(m_2,r_2)$. Therefore, the consistency of any posted ciphertext can be verified by re-encrypting with the claimed random value.
  \item \emph{Additive Homomorphism}: The multiplication of two ciphertexts is equal to the encryption of two corresponding plaintexts' addition, \ie,
      \begin{equation*}
            C_1 \times C_2 = E_n(m_1,r_1) \times E_n(m_2,r_2) = E_n(m_1+m_2,r_1\cdot r_2).
      \end{equation*}
\end{itemize}

In ERA, we guarantee the bid privacy and consistency by capitalizing the first two properties, whereas the homomorphic property is utilized to support auction verification.

\section{Privacy Preserving Integer Comparison}\label{sec:ppic}

In this section, we propose a privacy preserving integer comparison protocol, which is the basis of verification process.


In the privacy preserving integer comparison protocol, there are two major entities: a prover $\mathcal{P}$ and a verifier $\mathcal{V}$. The prover $\mathcal{P}$ knows two non-negative integers $x_1$ and $x_2$ as well as their comparison relation, \eg, $x_1 \geq x_2$. The main goal of the prover $\mathcal{P}$ is to convince the verifier $\mathcal{V}$ that the declared comparison relation is true without disclosing $x_1$ and $x_2$\footnote{Our privacy preserving integer comparison differs from the classical Yao's millionaires' problem~\cite{Yao:1982:PSC:1398511.1382751}, in which the prover $\mathcal{P}$ knows $x_1$ and the verifier $\mathcal{V}$ knows $x_2$.}. Here, we note that if $x_1 < x_2$ holds in fact, it is computationally infeasible for the prover $\mathcal{P}$ to generate a witness to $x_1 \geq x_2$.

For two non-negative integers $x_1,x_2 < n/2$, the inequality $x_1 \geq x_2$ holds if and only if $(x_1-x_2)~mod~n < n/2$. Thus, to demonstrate that $x_1 \geq x_2$, the prover $\mathcal{P}$ can equivalently show the following three inequalities hold:
\begin{equation}
x_1 < n/2,\quad x_2 < n/2,\quad (x_1-x_2)~mod~n < n/2.
\end{equation}
Here, we observe that the original integer comparison problem can be further reduced to a classical problem of range proof, \ie, proving in zero knowledge $x < n/2$. In particular, our range proof protocol is based on the bit representations of encrypted value. In addition, we prove $x < n/2$ by showing $x < 2^t < n/2$, where $2^t$ is a preset upper bound of $x$.

Before proposing the range proof protocol, we first introduce a test set $TS$, which is a set of $t$ Paillier-type ciphertexts:
\begin{align}
\nonumber
TS &= \left\{C_i| i \in [1, t]\right\} = \left\{E_n(m_i,r_i)| i \in [1, t]\right\}\\
   &= \left\{E_n(2^{i - 1},r_i)| i \in [1, t]\right\}.
\end{align}
We note that all elements in $TS$ should be ordered randomly to conceal the linkability between $C_i$ and $m_i$.

Given $C=E_n(x,r_x)$, the prover $\mathcal{P}$ can obliviously prove to the verifier $\mathcal{V}$ that $x<2^t<n/2$ by conducting the following two steps:

\noindent\textbf{Step 1: Proof Generation}

$x$ can be uniquely represented by a sum of distinct powers of 2:
\begin{equation}
x=2^{t_1}+2^{t_2}+\ldots+2^{t_k}.
\end{equation}
The prover $\mathcal{P}$ selects the ciphertext set $\mathbb{C}_x = \{C_{t_1},C_{t_2},\ldots,C_{t_k}\}$ of the plaintexts $2^{t_1},2^{t_2},\ldots,2^{t_k}$ from the test set $TS$, and derives a new random value $r^*$ from the corresponding random values $r_{t_1},r_{t_2},\ldots,r_{t_k}$ and $r_x$, where
\begin{equation}
r^*=(r_x^{-1}\times r_{t_1} \times r_{t_2} \cdots \times r_{t_k}) ~ (mod ~ n).
\end{equation}
The set of ciphertexts $\mathbb{C}_x$ and the random value $r^*$ are packaged as the proof, which is sent to the verifier $\mathcal{V}$.

\noindent\textbf{Step 2: Verification}

The verifier $\mathcal{V}$ verifies the correctness of the relation $x<2^t<n/2$ by checking whether
\begin{equation}\label{eq:check}
E_{n}^{-1}(x,r_x)\cdot C_{t_1} \cdot C_{t_2}\cdot\ldots\cdot C_{t_k} ~ (mod ~ n^2)=E_{n}(0,r^*).
\end{equation}
The above equation holds if and only if $x=2^{t_1}+2^{t_2}+\ldots+2^{t_k}$. Together with the fact that the cardinality of $\mathbb{C}_x$ is less than or equal to $t$, the verify $\mathcal{V}$ can conclude that
\begin{equation}
x = 2^{t_1}+2^{t_2}+\ldots+2^{t_k} <= 2^0 + 2^1 + \ldots + 2^ {t-1} < 2^t < n/2.
\end{equation}

Now, the prover $\mathcal{P}$ can convince the verifier $\mathcal{V}$ that $x_1 \geq x_2$ by applying the range proof protocol on the following three inequations:
$$
\left\{
\begin{aligned}
&x_1<2^t<n/2,\\
&x_2<2^t<n/2,\\
&(x_1-x_2)~ mod ~n<2^t<n/2.
\end{aligned}
\right.
$$

The correctness of the verification phase, especially Equation~(\ref{eq:check}), can be proved below:

\begin{proof*}
We first shift the term $E_{n}^{-1}(x,r_x)$ in Equation~(\ref{eq:check}) from the left hand side (LHS) to the right hand side (RHS), and get an equivalent form:
\begin{equation}\label{eq:check2}
C_{t_1} \cdot C_{t_2}\cdot\ldots\cdot C_{t_k} ~ (mod ~ n^2) = E_{n}(x,r_x) \cdot E_{n}(0,r^*) ~ (mod ~ n^2).
\end{equation}
Next, by applying the additive homomorphism of the Paillier encryption scheme, we expand the left and right hand sides of Equation~(\ref{eq:check2}), respectively:
\begin{align*}
\text{LHS}
&=  C_{t_1} \cdot C_{t_2} \cdot \ldots \cdot C_{t_k} (mod ~ n^2)\\
&= E_n(2^{t_1}, r_{t_1}) \cdot E_n(2^{t_2}, r_{t_2}) \cdot \ldots E_n(2^{t_k}, r_{t_k}) (mod ~ n^2)\\
&= E_n(2^{t_1}+2^{t_2}+\cdots+2^{t_k}, r_{t_1} \times r_{t_2} \ldots \times r_{t_k}) (mod ~ n^2),\\
\text{RHS}
&= E_n(x,r_x) \cdot E_n(0,r^*) ~ (mod ~ n^2)\\
&= E_n(x + 0, r_x \times r^*) ~ (mod ~ n^2)\\
&= E_n(x, r_x \times r_x^{-1} \times r_{t_1} \times r_{t_2} \ldots \times r_{t_k}) ~ (mod ~ n^2)\\
&= E_n(x, r_{t_1} \times r_{t_2} \ldots \times r_{t_k}) ~ (mod ~ n^2).
\end{align*}
Due to the uniqueness of Paillier-type ciphertext, we have
$$
\text{LHS} = \text{RHS}\quad\Leftrightarrow\quad x=2^{t_1}+2^{t_2}+\ldots+2^{t_k}.
$$
This completes our proof.\QEDA
\end{proof*}

\section{Design of ERA}\label{sec:ERA}

In this section, we propose ERA, which is an efficient, privacy-preserving, and verifiable online ad auction mechanism.

\subsection{Design Overview}\label{s5_1}

By exploiting the cryptographic tools in Section~\ref{sec:crypto} and the privacy preserving integer comparison protocol in Section~\ref{sec:ppic}, ERA achieves privacy preservation and verifiability simultaneously. In what follows, we illustrate the design challenges and the design rationales.




The first design challenge is privacy preservation in terms of both original bids and ad bidders' identities. We first consider bid protection. We introduce an agent to encrypt the original bids as mapped bids using the order preserving encryption scheme. Therefore, the ad networks and the auctioneer can learn the ranking order of bids by comparing the corresponding mapped bids to calculate the auction outcome. However, the semi-honest agent may reveal the original bids if she can obtain the mapped bids, and we tackle this vulnerability by taking two cooperative steps. First, each bidder fetches her/his mapped bid from the agent via oblivious transfer, which guarantees that the mapped bid selection does not leak her/his choice. Even so, the agent may still access the mapped bids on the certificated bulletin board. Thus, the auctioneer needs to provide one more encryption on the mapped bids before publication. In this way, as long as there exists no collusion between the agent and the auctioneer, the original bids can be well protected.


We then consider identity preservation, \ie, the bidding order in an internal auction can only be known by the managing ad network. Since we employ order preserving encryption to do the first-layer encryption, the order of mapped bids is exactly the same as that of original bids. If the identities of advertisers are not protected, the auctioneer may uniquely link a bidder with her/his bidding rank. To prevent this undesirable information leakage, each ad network needs to encrypt the identities of her/his bidder members
 The second design challenge comes from auction verification. The ad networks and the auctioneer in ad exchanges exclusively possess the bidding information, while any other system participant, as a verifier, cannot access it. This information asymmetry causes significant difficulties to our verifiable auction design. To solve this problem, we first introduce a certificated bulletin board to publish all the encrypted bidding information, including doubly encrypted bids and masked identities. We then employ the proposed privacy preserving integer comparison protocol to enable any verifier to check the order of {mapped} bids, \ie, the order of original bids, and thus verify the correctness of auction execution.


\subsection{Design Details}\label{s5_2}
We now introduce ERA in details. ERA consists of three stages: initialization, auction execution, and verification operation.

\subsubsection{Initialization}\label{s5_2_1}

The initialization stage contains two parts: bid and identity encryptions, and information publication.


\textbf{Bid and Identity Encryptions:} The original bid space $\Theta$ is defined as the set of $z$ possible bids:
$$\Theta =\left\{\theta_1,\theta_2,\ldots, \theta_z\right\},$$
where $\theta_1 \geq \theta_2 \geq \cdots \geq \theta_z$. Based on the original bid space, the agent runs the order preserving encryption scheme to generate a set of mapped bids:
$$\widehat{\Theta} =\left\{\hat{\theta}_1,\hat{\theta}_2,\ldots, \hat{\theta}_z\right\},$$
where $\hat{\theta}_i = OPES(\theta_i)$ and $\hat{\theta}_1 \geq \hat{\theta}_2 \geq \cdots \geq \hat{\theta}_z$. Without loss of generality, we assume that the maximal mapped bid is upper bounded by $2^t$ for some $t$, \eg, $t=32$ in our evaluation part.

Each bidder $s_i\in \mathbb{S}$ with original bid $b_i = \theta_{i'}$ contacts the agent to fetch her/his mapped bid $\hat{b}_i  = \hat{\theta}_{i'}$ from the mapped bid space $\hat{\Theta}$ via oblivious transfer. This guarantees that bidder $s_i$ only knows $\hat{\theta}_{i'}$, and has no idea of the other $z-1$ mapped bids in $\widehat{\Theta}$, while the agent does not know which mapped bid is chosen by the bidder $s_i$. However, the agent may still know the original bid of the bidder $s_i$ if she can access the mapped bid $\hat{b}_i$. Therefore, the ad network $a_j$, who is responsible for the bidder $s_i$, further encrypts the bid $\hat{b}_i$ using the Paillier scheme with the public key $n$ and a random value $r^1_i$. We note that the public key $n$ is provided by the auctioneer, and the random value $r^1_i$ is generated by the ad network $a_j$ for the bidder $s_i$. The doubly encrypted bid of the bidder $s_i$ is denoted by $c_i = E_n(\hat{b}_i, r^1_i)$.

For identity preservation, the ad network $a_j$ also needs to encrypt the identities of her/his subordinate bidders. The ad network $a_j$ encrypts the identity of bidder $s_i$ by adopting the Paillier scheme using the public key $n_j$ and the random value $r^2_i$, which are both generated by the ad network $a_j$. The masked identity of the bidder $s_i \in \mathbb{S}$ is denoted by $id_i=E_{n_j}(s_i, r^2_i)$.


\textbf{Information Publication:} To facilitate the auction verification, the following information should be posted on the certificated bulletin board:
\begin{itemize}
\item $l-1$ test sets $\{TS_1, TS_2, \cdots, TS_{l-1} \}$: these test sets are posted by the auctioneer with her/his signature, and will be used to verify the comparison relation of the $l$ bids.
\item $l$ commitments $\{COM_1, COM_2, \cdots, COM_l\}$: the commitment of the bidder $s_i\in \mathbb{S}$ is defined as $COM_i = (c_i, id_i)$. These commitments are calculated and posted by all the ad networks, and will be used to verify the auction outcome.
\end{itemize}

\subsubsection{Auction Execution}\label{s5_2_2}
The auction execution is divided into two stages: the internal auction stage and the global auction stage.
Each ad network $a_j\in \mathbb{A}$ conducts an internal auction among her/his bidder members. The ad network $a_j$ selects the highest and the second highest mapped bids, both of which are sent to the auctioneer with the signature by $a_j$.
 In the global auction stage, the auctioneer chooses the globally highest and second highest mapped bids, \ie, $\hat{b}_{max}$ and $\hat{b}_{sec}$, from the internal auction outcomes provided by all the ad networks. Finally, the auctioneer determines the identity of winner $s_{max}$ and corresponding payment $b_{sec}$: the auctioneer first delivers the highest mapped bid $\hat{b}_{max}$ to the ad network who has submitted it, and the ad network replies with the identity of the winner $s_{max}$; the auctioneer then sends the second highest mapped bid $\hat{b}_{sec}$ to the agent, and the agent feeds back the winner's payment $b_{sec}$, which is the original bid of $\hat{b}_{sec}$. We note the agent can obtain this payment $b_{sec}$ by using the inverse function of $OPES(\cdot)$, \ie, $b_{sec}=OPES^{-1}(\hat{b}_{sec}).$

\subsubsection{Verification Operation}\label{s5_2_3}

After the auction execution stage, the ad networks in charge of the bidders with $\hat{b}_{max}$ and $\hat{b}_{sec}$ are required to mark their commitments on the certificated bulletin board for verification. We denote these two marked commitments as $COM^{*}_{max} = (c^*_{max}, id^*_{max})$  and $COM^{*}_{sec} = (c^*_{sec}, id^*_{sec})$.


According the definition of second-price auction~\cite{vickrey1961counterspeculation2}, the auction outcome is correct means that the ad space is sold to the bidder with the globally highest bid $b_{max}$, and her/his payment equals to the globally second highest bid $b_{sec}$. Formally, we claim that the auction outcome is correct if the following two conditions are satisfied:

$$
\left\{
\begin{aligned}
&b_{max}\geq b_{sec},\\
&b_{sec}\geq b_i, \forall i\neq max.
\end{aligned}
\right.
$$

During the verification phase, we assume that the auctioneer serves as the prover $\mathcal{P}$, and any party can decide to be a verifier $\mathcal{V}$. We now describe the verification process, which consists of three components: winner and payment verifications, ordering verification, and patching verification.

\noindent \textbf{Step 1: Winner and Payment Verifications}

In this step, the verifier $\mathcal{V}$ wants to check whether the declared outcome $(s_{max},b_{sec})$ is consistent with the marked outcome $(s^*_{max}, b^*_{sec})$.
This verification is based on the computationally binding property of the Paillier homomorphic encryption scheme.

First, for winner verification, the ad network $a_j$, who is responsible for the winner $s_{max}$,  provides her/his public key $n_j$ and the random value $r^2_{max}$. The verifier $\mathcal{V}$ can verify the correctness of the winner's identity by checking whether the re-encrypted identity $id_{max} = E_{n_j}(s_{max}, r^2_{max})$ is the same as the marked identity on the certificated bulletin board, \ie, $id_{max} = id^*_{max}$.

Second, regarding payment verification, the auctioneer sends the public key $n$ and the random value $r^1_{sec}$\footnote{Using private key $\phi$, the auctioneer can recover the random value $r^1_{sec}$, which is generated by the ad network} to the verifier $\mathcal{V}$. In order to check the consistency of the declared payment, the verifier $\mathcal{V}$ needs to ask the agent for the mapped payment $\hat{b}_{sec}$ by revealing the the original payment $b_{sec}$. Next, the verifier $\mathcal{V}$ re-encrypts $\hat{b}_{sec}$ as $c_{sec} = E_n(\hat{b}_{sec},r^1_{sec})$. The verifier $\mathcal{V}$ then examines whether $c_{sec}$ is equal to the marked doubly encrypted payment on the certificated bulletin board, \ie, $c_{sec} = c^*_{sec}$.

The failure of winner and payment verifications implies that either the ad exchange misreported the outcome or the ad networks marked the wrong commitments. Under such circumstance, the ad exchange needs to work with the corresponding ad networks until the winner and payment verifications pass. After that, Step 2 begins.



\noindent \textbf{Step 2: Ordering Verification}
\begin{figure}[!t]
    \centering
    \includegraphics[width = 0.95\columnwidth]{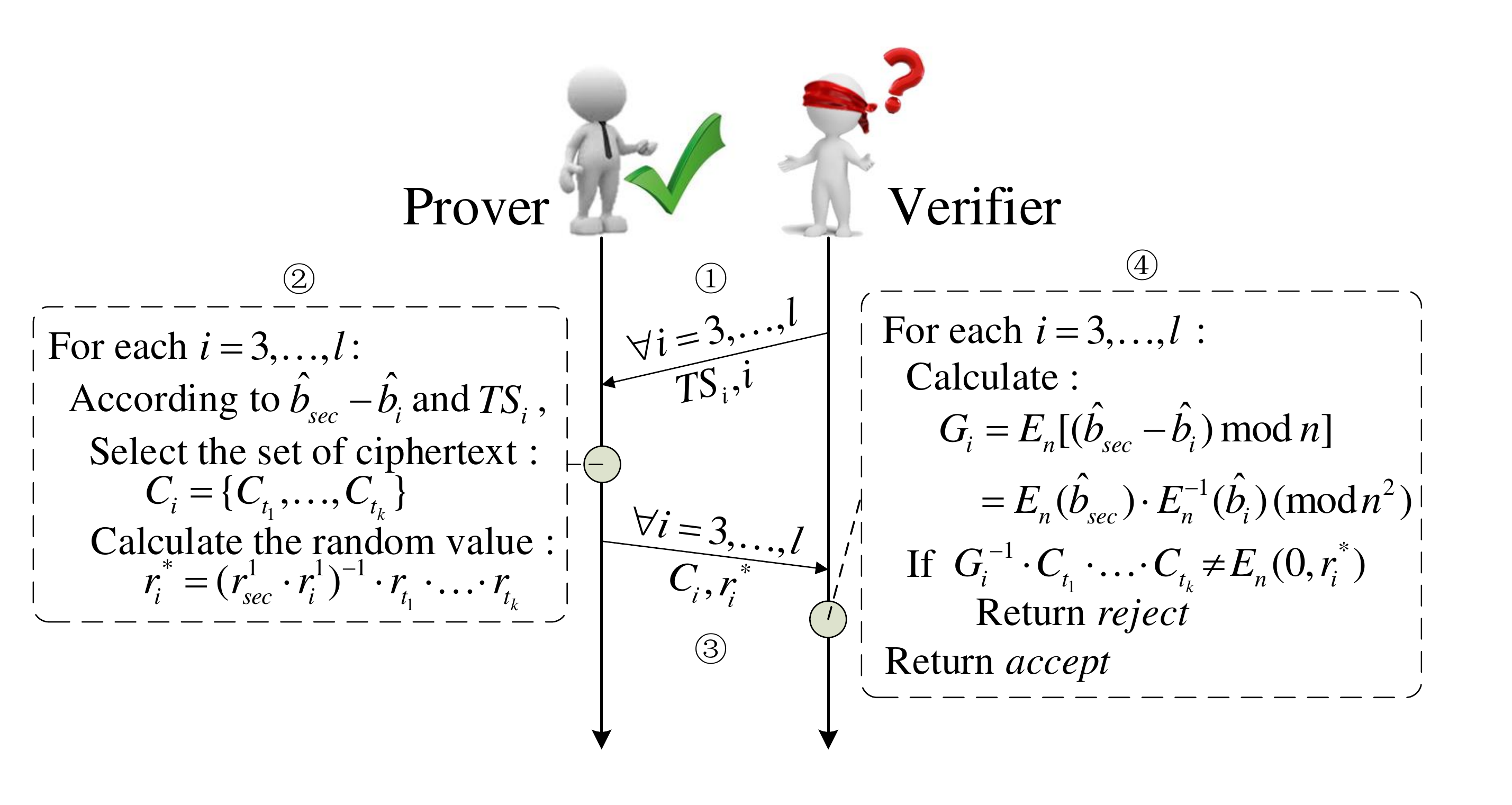}
    \caption{The process of ordering verification.}
    \label{fig:2}
\end{figure}


As the guarantee of order preserving encryption, we can examine the order of mapped bids to see whether the order of original bids is correct. We assume that the mapped bids are sorted in a non-increasing order:
$$
\Gamma: \hat{b}_1 \geq \hat{b}_2 \geq \cdots \geq \hat{b}_l,
$$
where $\hat{b}_1 =  \hat{b}_{max}$ and $\hat{b}_2 =  \hat{b}_{sec}$. For ordering verification, the prover $\mathcal{P}$ should prove that the mapped payment $\hat{b}_{sec}$ is equal to or less than the winner's {mapped} bid $\hat{b}_{max}$, and  $\hat{b}_{sec}$ is equal to or greater than the other $l-2$ mapped bids.
Since the $l$ {mapped} bids provided by the agent are all in the range $[1, 2^t]$, $2^t \leq n/2$, the correctness of the {mapped} bids ordering $\Gamma$ can be verified by applying the privacy preserving integer comparison protocol in Section~\ref{sec:ppic} over the $l-1$ pairwise comparisons, \ie, $\left\langle\hat{b}_{sec}, \hat{b}_i \right\rangle,\forall \ i \neq sec$.

We depict the verification for the $l-2$ comparisons $\left\langle\hat{b}_{sec}, \hat{b}_i \right\rangle,\forall \ 3 \leq i \leq l $ in Figure~\ref{fig:2}. The relation between $\hat{b}_{max}$ and $\hat{b}_{sec}$ can be verified in a similar way. In order to verify the relation $\left\langle\hat{b}_{sec}, \hat{b}_i \right\rangle$, the verifier $\mathcal{V}$ chooses a certain test set $TS_{i}$, and sends it with the index $i$ to the prover $\mathcal{P}$. The prover $\mathcal{P}$ then constructs the ciphertext set $\mathbb{C}_i = \left\{C_{t_1}, C_{t_2} \cdots, C_{t_k}\right\}$ such that $\hat{b}_{sec} - \hat{b}_i  = 2^{t_1} + 2^{t_2} + \cdots + 2^{t_k}$, and calculates a new random value $r^*_i$. Both the set $\mathbb{C}_i$ and the random value $r_i^*$ are sent back to the verifier $\mathcal{V}$, who then calculates $G^{-1}_i \times C_{t_1} \times \cdots \times C_{t_k}$ and $E_{n}(0, r^*_i)$ to determine whether to accept or to reject the ordering verification.



\noindent \textbf{Step 3: Patching Verification}

If the first two verification steps do not pass, the auctioneer is accused of cheating, unless she can provide the evidence that the fault of the outcome is caused by some ad networks. The auctioneer uses her/his private key $\phi$ to decrypt the doubly encrypted bids on the certificated bulletin board to obtain all mapped bids. Then, the auctioneer resorts these mapped bids to check the correctness of the internal auction outcome submitted by each ad network, and thus seeks out those misbehaved ad networks.


\subsection{Security Analysis}\label{s5_3}
In this section, we analyze the security of ERA.

\begin{theorem}
ERA guarantees the property of privacy preservation in terms of bids and identities.
\end{theorem}
\begin{proof*}
We first consider bid protection. We note that the original bids in ERA are doubly encrypted, first by order preserving encryption and then by Paillier encryption, before being posted on the certificated bulletin board. Moreover, order preserving encryption is secure under ciphertext-only attack~\cite{Agrawal:2004:OPE:1007568.1007632}, and Paillier encryption is semantically secure~\cite{PP99}. Thus, only the party, who owns the private keys of these two encryption schemes simultaneously, can reveal the original bids. However, these two private keys are separately kept by two different parties in ERA, where the agent holds the private key for the order preserving encryption scheme, while the auctioneer holds the private key for the Paillier homomorphic encryption scheme. Since we have assumed that there exists no collusion between the agent and the auctioneer, none but each bidder can know her/his original bid, and thus the bid protection is achieved.

We then consider identity preservation. In ERA, each bidder's identity is encrypted by her/his ad network using the Paillier encryption scheme, which provides semantic security. By definition, except the managing ad network, any probabilistic polynomial-time adversary cannot distinguish the original identities of different ad bidders. Thus, the auctioneer cannot link the bid ranking with a specific bidder.

In conclusion, ERA achieves the privacy preservation of advertisers.
\QEDA
\end{proof*}



\begin{theorem}
ERA is a verifiable ad auction mechanism.
\end{theorem}
\begin{proof*}
We claim that if both Step 1 and Step 2 in the verification operation stage pass, the outcome of the auction is provably correct.
If Step 1 passes, due to the uniqueness of Paillier-type ciphertext, the winner and corresponding payment published by the ad exchange are exactly consistent with the marked ones on the certificate bulletin board, \emph{\ie}, $\hat{b}_{max}=\hat{b}^*_{max}$ and $\hat{b}_{sec}=\hat{b}^*_{sec}$. If Step 2 also passes, on one hand, the bid of the marked winner is greater than or equal to the marked payment, \emph{\ie}, $\hat{b}^*_{max}\geq \hat{b}^*_{sec}$; on the other hand, the marked payment is equal to or greater than the other $l-2$ mapped bids, \emph{\ie}, $\hat{b}_{sec}^*\geq \hat{b}_i, \forall i\neq sec,max$. Hence, the rules of second-price auction are satisfied, and ERA is a verifiable ad auction mechanism.  \QEDA
\end{proof*}

\section{Evaluation Results}\label{sec:eval}

ERA integrates several cryptographic tools to guarantee privacy preservation and verifiability. A practical ad auction mechanism should incur low cost in terms of computation, communication, and storage overheads, such that it can be deployed in today's ad exchanges.
We show the evaluation results of ERA in this section.




\textbf{Simulation Setting: }
We have implemented ERA using network simulation.
 The possible original bids range from \$0.01 to \$100 with an increment of \$0.01, which is typically the smallest billable unit in today's ad exchanges~\cite{RTBPB59}.
 The maximal mapped bid is set as $2^{32}$. In the oblivious transfer, the group order $\rho$ is 1024-bit long, and the size of $\xi_i$ is bounded in 32 bits. The Paillier homomorphic encryption scheme is implemented using an open-source library~\cite{AdvancedCryptoSoftwareCollection}, in which the public key is set 1024-bit long. The running environment is a standard $64$-bit Ubuntu $14.04$ Linux operating system on a desktop with Intel(R) Core(TM) $i5$  $3.10GHz$.

\subsection{Computation Overhead}\label{s6_1}
We now show the computation overhead of three important components in ERA, \ie, mapped bid generation, auction execution, and verification.

\subsubsection{Mapped Bid Generation}\label{s6_1_1}
\begin{figure}[t]
    \centering
    \includegraphics[width=0.95\columnwidth]{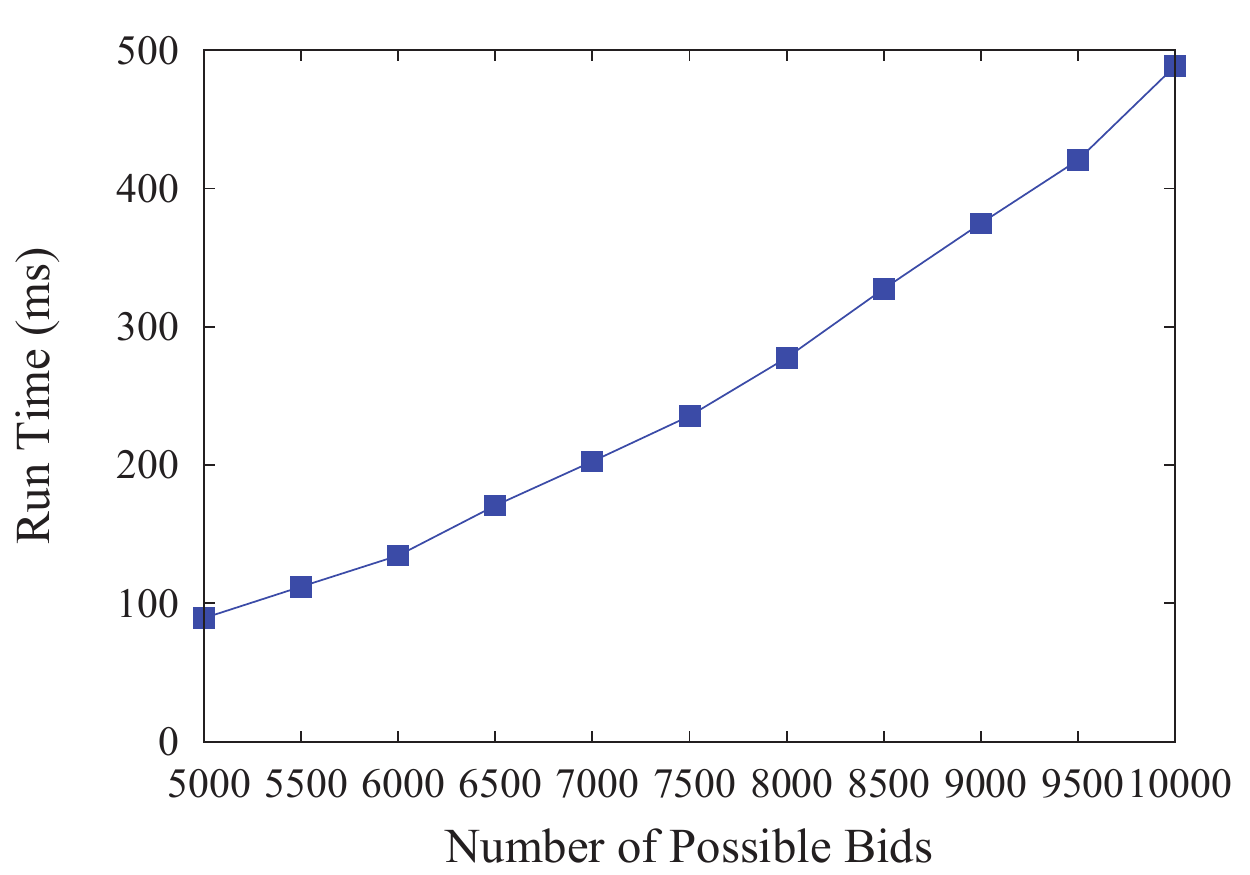} 
    \caption{Computation overhead of mapped bid generation for each bidder.}
    \label{fig:3}
\end{figure}

By averaging 1000 simulation instances, in Figure~\ref{fig:3}, we plot the computation overhead of the agent for generating mapped bid for each bidder, when the number of possible bids increases from 5000 to 10000 with a step of 500. We can see that the computation overhead grows linearly with the number of possible bids, and reaches around 500ms at 10000 possible bids. This is because the computation overhead mainly comes from running the oblivious transfer, in which the agent should calculate $z$ auxiliary messages, \ie, $\{\xi_i| 1 \leq i \leq z\}$. This implies that the computation overhead of mapped bid generation is proportional to the number of possible bids $z$.

\subsubsection{Auction Execution}\label{s6_1_2}
\begin{figure}[!t]
\centering
\includegraphics[width = 0.95\columnwidth]{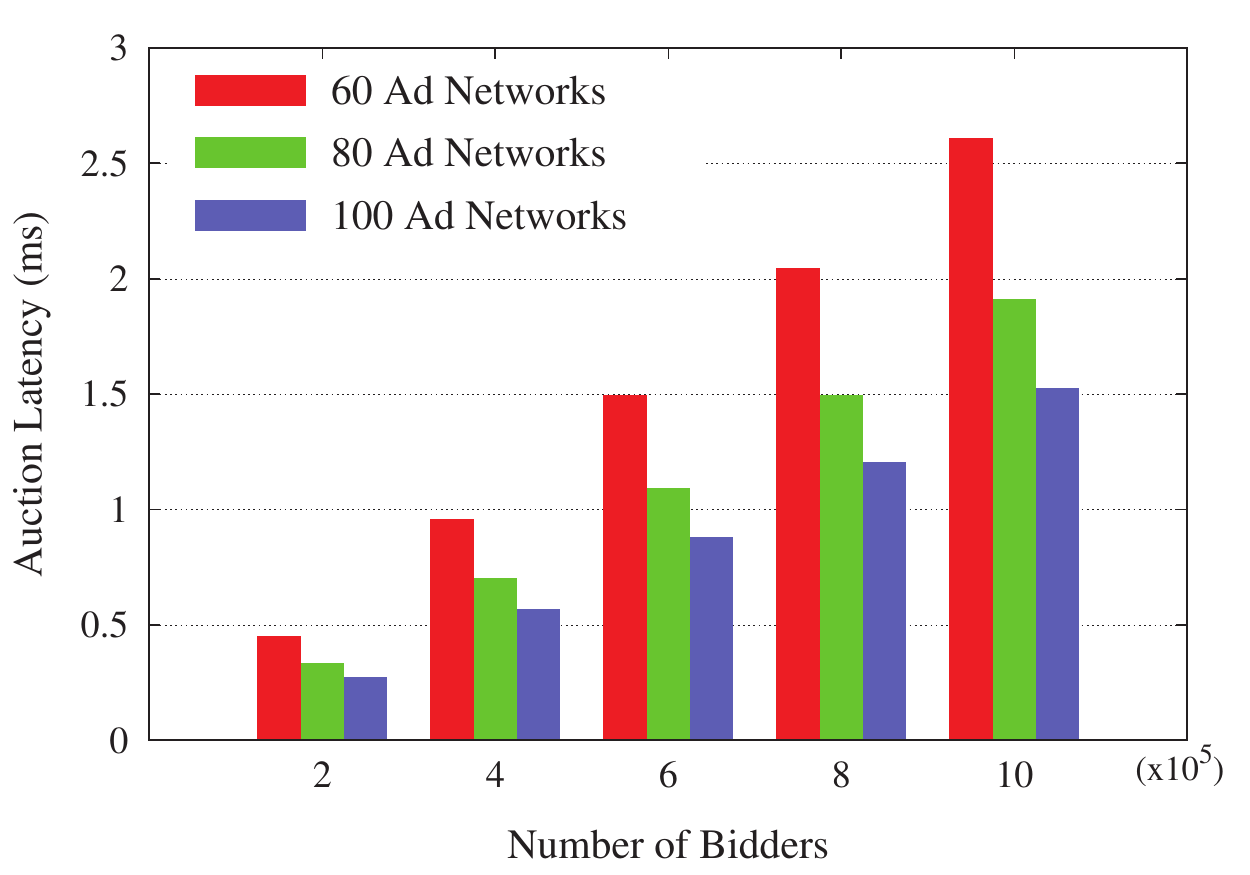}
\caption{Auction latency of ERA.}
\label{fig:4}
\end{figure}

We measure the metrics of auction latency and auction scale to understand the computation overhead of the auction execution in ERA.

Figure~\ref{fig:4} shows the auction latency of ERA when the number of bidders ranges from $2\times 10^5$ to $10\times 10^5$ with an increment of $2\times 10^5$, and the number of ad networks can be chosen as $60$, $80$, and $100$. The first observation from Figure~\ref{fig:4} is that when the number of ad networks is fixed, the auction latency of ERA increases with the number of bidders. The second key observation is that when the number of bidders is fixed, the auction latency decreases with the number of ad network.

\begin{table}[!t]
\caption{\textsc{Auction Latency (ms).}} \label{tab:2}
\centering
\resizebox{\columnwidth}{!}{
\begin{tabular}[t]{c|ccccc}
\hline
\#bidders ($\times10^4$) & 2 & 4 & 6 & 8 & 10\\
\hline
\hline
ERA with 100 ad networks & 0.27 & 0.57 & 0.88 & 1.21 & 1.53\\
\hline
Benchmark model & 39.47 & 82.73 & 126.85 & 171.32 & 212.56\\
\hline
\end{tabular}
}
\end{table}

For more intuitive comparison, we also list the auction latency of ERA with 100 ad networks and that of the benchmark model with none ad network (\ie, traditional second-price auction model) in Table~\ref{tab:2}. We can see that in a small-scale auction, \eg, $2\times 10^5$ bidders, ERA is roughly $50\times$ better than the {benchmark} model. When the scale of auction becomes larger, ERA's advantage over the {benchmark} model is more remarkable. Furthermore, we set the upper bound of the auction latency to be 10ms, and thus compare the maximal bidders that ERA and the benchmark model can support, respectively. The evaluation results show that the {benchmark} model can only handle up to $57,200$ bidders, while ERA can support $5,704,200$ bidders.

These evaluation results demonstrate that ERA can indeed help to reduce the auction latency by introducing a proper number of intermediary ad networks, especially in large-scale ad auctions.


\subsubsection{Verification}\label{s6_1_3}

\begin{table}[!t]
\caption{\textsc{Computation overhead of verification (s).}} \label{tab:1}
\centering
\resizebox{\columnwidth}{!}{
\begin{tabular}[t]{c|m{3cm}<{\centering}m{2.6cm}<{\centering}m{2.6cm}<{\centering}m{2.6cm}<{\centering}}
\hline
         & \multicolumn{2}{c}{Preparation} & \multicolumn{2}{c}{Operation}\\
\hline
\hline
\#bidders ($\times10^4$)  & Test Set Generation & Commitment Generation &  Ordering & Patching\\
\hline
2  & 1.00$\times10^4$    & 316.85 & 132.26 & 117.08\\
4  & 2.00$\times10^4$    & 631.52  & 265.69 & 233.58\\
6 & 3.00$\times10^4$     & 946.46  & 393.86 & 349.78\\
8 & 4.00$\times10^4$     & 1263.91  & 527.03 & 466.04\\
10 & 5.00$\times10^4$    & 1577.34  & 656.84 & 582.61\\
\hline
\end{tabular}
}
\end{table}

We now investigate the computation overhead of the verification, including the preparation and operation phases. In this set of simulation, we set the maximum number of bidders and the number of ad networks to be $10^5$ and $100$, respectively.
Table~\ref{tab:1} lists the evaluation results.

The preparation phase can be further divided into two parts: the generation of test sets by the auctioneer and the generation of commitments by ad networks. From Table~\ref{tab:1}, we can find that the auctioneer has higher computation overhead (about $31.70\times$) than that of each ad network. This outcome stems from that the auctioneer should calculate $l-1$ test sets, around 0.5s for each test set generation. In contrast, each ad network just needs to generate commitments for her/his bidder members.

The computation overhead of the operation phase mostly comes from the ordering verification and patching verification\footnote{The computation overhead of the winner and payment verifications is omitted here because it is extremely lower than the other two steps.}.
From Table~\ref{tab:1}, we can see that the computation overheads of these two components increase with the auction scale. In particular, when the number of bidders reaches $10^5$, the time overheads of ordering verification and patching verification are $656.84$s and $582.61$s, respectively. Combining with Figure~\ref{fig:4}, we can analyze that the computation overhead of verification is higher than that of auction execution. Since the verification can be conducted off-line, the running time constrain on verification is not so strict. Hence, the time consumption of verification is affordable when ERA is integrated into practical ad exchanges.

\subsection{Storage and Communication Overheads}\label{s6_2}

\begin{figure}[t]
\centering
\includegraphics[width = 0.95\columnwidth]{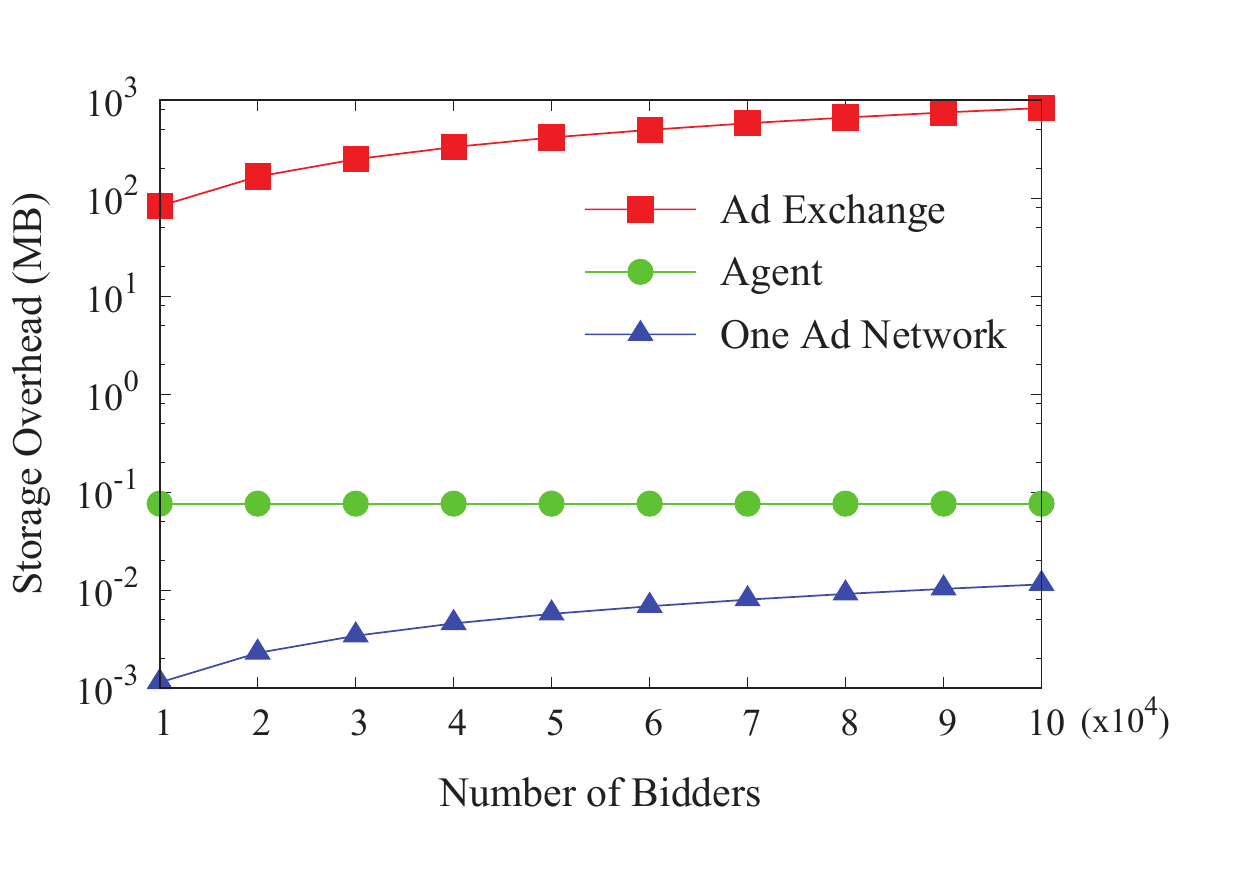}
\caption{Storage overhead of ERA.}\label{fig:6}
\end{figure}

Figure~\ref{fig:6} plots the storage overheads of the ad exchange, the agent, and each ad network, where the number of possible bids is fixed at $10000$, the maximal number of bidders is set to be $10^5$, and the number of ad networks is $100$.
 We can see that the storage overheads of the ad exchange and each ad network grow linearly with the number of bidders, while the storage overhead of the agent remains unchanged. The reason is that the storage overhead of the agent mostly comes from storing the original and mapped bids, which is independent of the number of bidders. We can also find that the ad exchange spends much more storage space than each ad network. This is because the ad exchange needs to maintain the certificated bulletin board, on which all test sets and commitments are posted, whereas each ad network only stores the mapped bids, identities, and ad tags of her/his bidder members.

We also measure ERA's communication overhead, which is mainly incurred by the interactions in oblivious transfer and ordering verification. On one hand, in the oblivious transfer, each bidder receives 10000 32-bit long messages, \ie, $\{\xi_i|i \in [1, 10000]\}$. On the other hand, the prover and the verifier need to transfer $\{TS_i, i, C_i, r_i^* | i \in [1, l], i \neq 2\}$ in the ordering verification, which speeds $800$MB bandwidth when the number of bidder is $10^5$.



\section{Related Work}\label{sec:related:work}

In this section, we briefly review the related work about the privacy preserving and verifiable auction design.

Inspired by early works~\cite{NS93,398918}, various privacy preserving and verifiable auction mechanisms have been extensively studied. The existing works generally fall into the following three categories with different auction models.

 \textbf{No Auctioneer}: Bidders themselves jointly determine the auction outcome by adopting the ideas from secure multiparty computation ~\cite{Yao:1982:PSC:1398511.1382751,Brandt:2008:EUP:1330332.1330338,Chaum:1988:MUS:62212.62214}. These mechanisms ensure bid protection and auction correctness, but induce unaffordable computation and communication overheads. Therefore, these mechanisms are inefficient and impractical in the scenario of ad exchanges.

 \textbf{One Auctioneer}:  One auctioneer is responsible for conducting the full auction. Parkes~\et~\cite{parkes2008practical} proposed a method based on the Paillier cryptosystem to achieve verifiability. Inspired by secret sharing, Rabin~\et~\cite{RMMY12,rabin2007highly} developed an efficient, novel, and secure solution for validating the correctness of an auction outcome. However, their method requires the leakage of bidding information to the auctioneer.

 \textbf{Additional Third Party}: A third party is introduced to cooperate with the auctioneer to run the auctions. The scheme proposed by Naor \et~in~\cite{Naor:1999:PPA:336992.337028} constructed a boolean circuit, which calculates the auction outcome for any given set of bids. Based on RSA, Juels and Szydlo proposed a privacy preserving auction mechanism with a reasonable computational complexity~\cite{US03}. Specific to online applications, a number of auction mechanisms~\cite{6566870,huang2015pps,huang2015general} get a good tradeoff between performance and security. Unfortunately, these works only consider bid protection, but ignore the problem of verification. ERA belongs to this category, and moves forwards to guarantee verifiability and privacy preservation at the same time.

When these existing privacy preserving and/or verifiable auction mechanisms are directly applied to ad exchanges, they have high computation and communication complexity. Thus, these works could only support auctions with small scale~\cite{Brandt:2008:EUP:1330332.1330338,BF06,jung2013efficient} or a limited number of possible bids~\cite{Lipmaa:2002:SVA:1765278.1765285,dreier2015brandt,howlader2015sealed}, which may be unacceptable for today's ad exchanges.

The most relevant work is paper~\cite{angel2013verifiable}, in which an online verifiable auction mechanism for ad exchanges was proposed. However, the bid protection was not considered, and the interactions between the ad exchange and the ad bidders are too frequent. Therefore, jointly considering both security and efficiency requirements, ERA is the first efficient, privacy preserving, and verifiable online auction mechanism for ad exchanges.
\section{CONCLUSION}\label{sec:conclusion}
In this paper, we have proposed the first secure mechanism ERA for ad exchanges, achieving both privacy preservation and verifiability. In ERA, the outcomes of ad auctions can be calculated and verified to be correct, while maintaining the private information of advertisers. We have implemented ERA, and extensively evaluated its performance. Evaluation results have demonstrated that ERA satisfies the desirable properties of low-latency and large-scale for practical ad exchanges.

As for future work, one possible direction is to employ a stronger but still time-efficient encryption scheme to achieve privacy preservation during auction calculation and verification. Another interesting direction is to consider the security issues in the allocation of packaged ad spaces.





\section*{REFERENCE}

\bibliographystyle{elsarticle-num}
\bibliography{ERA}



\end{document}